\documentstyle[aps,prl,multicol,epsf,epsfig,float]{revtex}

\newcommand{\ri}{{\rm i}}
\newcommand{\re}{{\rm e}}

\title{Quantum Chaos and Quantum Algorithms}
\author{Daniel Braun}
\address{c/o
IBM T.J. Watson Research Center, Yorktown Heights, NY 10598}

\begin{document}

\maketitle
\draft

\begin{abstract}
\begin{center}
\parbox{14cm}{It was recently shown (quant-ph/9909074) that parasitic
random interactions 
between the qubits in a quantum computer can induce quantum chaos and
put into question the operability of a quantum computer. In this work I
investigate whether already the interactions between the qubits introduced with
the intention 
to operate the quantum computer may lead to quantum chaos. The analysis
focuses on two well--known quantum 
algorithms, namely Grover's search algorithm and the quantum Fourier
transform. I 
show that in both cases the same very unusual combination of
signatures from chaotic and from integrable dynamics arises.}
\end{center}
\end{abstract}

\vspace*{0.2cm}
\noindent

\begin{multicols}{2}

The problem of quantum chaos in quantum computers (QC) \cite{Deutsch85} has
recently 
attracted considerable attention after a pioneering work by  
Georgeot and Shepelyansky \cite{Georgeot99}. These authors pointed out
that residual, uncontrolled interactions between qubits might induce
quantum chaos in the QC if the   
strength of the interaction exceeds a certain critical level, and they
argued that this might 
destroy the operability of the QC. While the model considered by Georgeot
and Shepelyansky \cite{Georgeot99} does not describe a particular physical
realization of a QC, and in particular does not allow for time dependent
operating of quantum gates, it is sufficiently generic to mimic a quantum
register, i.e.~a (static) memory of the QC, in which a state can be stored
and from which it should be retrievable again at a later time. 

It is clear that residual interactions between the qubits will in general
lead to 
eigenstates of the quantum register that are not the original 
product states $|0\rangle\ldots|0\rangle|0\rangle$,
$|0\rangle\ldots|0\rangle|1\rangle$, 
$\ldots$ $|1\rangle\ldots|1\rangle|1\rangle$ (called multi--qubit states in the
following). Therefore,  
information stored in the register will evolve 
with 
time. If the exact eigenstates of the register are superpositions  
of only a few multi--qubit states, the register will oscillate 
quasiperiodically   between these. However, in the case of quantum chaos,
the eigenstates of the register are superpositions  of practically {\em all}
$2^n$ multi--qubit states. 
The quasiperiodic oscillations due to the interaction between qubits degradate
then for  all practical times to 
a decay of the original re\-gis\-ter state into all multi--qubit states. The
time scale $\tau_\chi$ for this decay is set by the inverse width of the
distribution of 
eigenenergies. Georgeot and Shepelyansky concluded that all calculations of
the quantum  
computer should have finished long before the time $\tau_\chi$ and that this
would limit the operability of the computer in a very similar fashion as
decoherence. The critical interaction strength $J_c$ between qubits at which
quantum chaos sets in decreases like $J_c\propto 1/n$ with the 
number $n$ of qubits, and the transition to quantum chaos becomes more and
more abrupt with increasing $n$ \cite{Georgeot99}. 

Later, Silvestrov et al.~questioned Georgeot's and Shepelyansky's
conclusions \cite{Silvestrov00}. They showed that even in the case of
quantum chaos error 
correction schemes \cite{Shor95,Steane96} are  capable of dealing with
errors generated by 
quantum chaos. However, much more error correction is needed than in the
absence of quantum chaos. Their model again included only static
interactions between qubits.  

In this work I examine a question that is somewhat complementary to the one
investigated in \cite{Georgeot99,Silvestrov00}: Is it possible that already the
interactions 
introduced on purpose between qubits in order to operate the quantum
computer lead to quantum chaos, even if there are no
residual parasitic interactions  
between qubits? This is an intriguing question, since it has implications
for the amount of resources necessary for implementing a given
algorithm. 
Already classically chaotic algorithms (e.g. for calculating the time
evolution of a classical chaotic system) require much more computing
resources than integrable ones: Since errors in the initial conditions
amplify exponentially in time, one needs to spend many more bits than for
calculating an integrable time evolution over the same time.  

In quantum mechanics chaos manifests itself by a sensitivity not to the
initial state but to the {\em control 
parameters} \cite{Peres91} (amongst other signatures, see below). The
fidelity $|\langle 
\psi(k)|\tilde{\psi}(k)\rangle|^2$ of a 
wave--function $|\psi(k)\rangle=U^k(\lambda)|\psi\rangle$  with
respect to a  
wave--function $|\tilde{\psi}(k)\rangle=U^k(\tilde{\lambda})|\psi\rangle$
decreases exponentially with the 
discrete time $k$, if
$\lambda$ and $\tilde{\lambda}$ are two slightly different system control
parameters and the time evolution $U(\lambda)$ is chaotic. In the
case of integrable quantum dynamics 
the fidelity typically shows quasiperiodic oscillations in $k$. Therefore a
chaotic quantum algorithm will need more resources in the form of more
precise quantum gates, or as pointed out in \cite{Silvestrov00}, more error
correction. 

A quantum algorithm (QA) is uniquely defined by the unitary
transformation $U$ it induces in the entire multi-qubit Hilbert space, and
the question of quantum chaos can be studied directly on the level of that
unitary transformation, without the need to deal with the time dependent
Hamiltonian by which it is generated. This is what I am going to do in this
work, therefore being able to go beyond the models with
time-independent Hamiltonians studied in \cite{Georgeot99,Silvestrov00}. 

Obviously, the answer to the question posed depends on the quantum
algorithm. 
A quantum algorithm simulating a
quantum chaotic system 
\cite{Song01,Georgeot01} is by definition a unitary transformation showing
quantum 
chaos, and will thus need very precise tuning of the control
parameters. In this paper I focus, however, on two of the most well--known
quantum 
algorithms,  where the answer is less clear from the beginning, namely
Grover's search  
algorithm \cite{Grover97} and the quantum Fourier transform (QFT). The
latter is the
center piece of several important QAs, like phase estimation, order-finding,
the hidden subgroup problem (see \cite{Nielsen00}), and, 
most prominently, Shor's factoring
algorithm  \cite{Shor94}. Also from
the pure quantum chaos point of view the question of quantum chaos in
these algorithms is very interesting.  In fact it turns out that both
algorithms have symmetry properties that lead to a remarkable and 
very non generic mixture of signatures of quantum chaos and quantum
in\-te\-gra\-bi\-li\-ty. 

The first thing that comes into mind for checking for quantum chaos, is
 the eigenvalue 
 and eigenvector statistics of $U$:
 It is believed that an eigenvalue-- and eigenvector
statistics of $U$ corresponding to Dyson's circular ensembles indicates
quantum chaos \cite{Bohigas84,Berry84}. That is, the
eigenvalues $\lambda_i=\exp(-\ri\varphi_i)$ should show universal level
repulsion \cite{proof}.  In
the limit of large $N$ one expects a distribution $P(s)$ of nearest neighbor
spacings $s_i=N(\varphi_{i+1}-\varphi_i)/2\pi$ that is well described by
the universal Wigner--Dyson statistics \cite{Mehta91}. If $U$ is
covariant under any anti--unitary operation $T$ that squares to unity there is
always a basis in which the eigenvectors of $U$ can be chosen real. The
relevant random matrix ensemble is 
then the  circular orthogonal ensemble (COE) with a $P(s)$ 
very well approximated by
\begin{equation}\label{WD}
P(s)=\frac{s\pi}{2}\re^{-s^2\pi/4}\,.
\end{equation}
The best known example is conventional time reversal symmetry, in which case
$T$ is the complex conjugation operator \cite{Haake91}.
The eigenvector statistics is usually described in terms of a distribution
of eigenvector components. Picking any component $c_{i}$ of any eigenvector at
random, random matrix theory (RMT) predicts for $y=N|c_{i}|^2$ in the COE
case the 
so--called Porter--Thomas distribution,
\begin{equation} \label{PT}
R_{\rm COE}(y)=\frac{1}{\sqrt{2\pi y}}\re^{-y/2}\,.
\end{equation}

Let us now have a look at Grover's algorithm \cite{Grover97}. It allows to
find an entry with index $\xi$ in a unsorted quantum database that is
distinguished from the 
others by a given property. The distinction may be formalized by an oracle
query $O$ which in the multi--qubit basis is a unitary
diagonal matrix with entries $O_{ii}=1-2\delta_{i\xi}$ where
$\delta_{i\xi}$ is the Kronecker delta. Thus, presented a
register state the oracle
always gives back the same state unless it is the searched one in which case
the oracle changes the state's phase by $\pi$. Grover's algorithm commences
with the
Hadamard transformation 
\begin{equation} \label{hada}
H=H_0\otimes H_2\otimes\ldots\otimes H_{n-1}\,,
\end{equation}
where 
\begin{equation} \label{hi}
H_i=\frac{1}{\sqrt{2}}\left(
\begin{array}{cc}
1&1\\
1&-1
\end{array}
\right)
\end{equation}
is a Hadamard transformation for the $i$th qubit. Starting from the
register state $|0\rangle \ldots |0\rangle$, the Hadamard transformation
brings the system into a superposition of all multi--qubit states with equal
weight $1/\sqrt{N}$. Next comes an iteration of the oracle query followed 
by a ``diffusion matrix'' $D$. The latter has matrix elements 
$D_{ij}=2/N-\delta_{ij}$.
The total algorithm thus reads
\begin{equation} \label{Grover}
U_G=(DO)^pH\,.
\end{equation}
The optimal value for the integer $p$ is given by $p=[\pi/(4\theta)]$ with
$\sin^2\theta=1/N$ ($[.]$ denotes the integer value)
\cite{Boyer96}, and I have chosen this value for all
calculations. Remarkably, the quantum  
computer finds the searched element with $\sim \sqrt{N}$ queries, whereas a
classical computer when presented with the same problem would have to ask
the oracle on the average $N/2$ times.

From the definition of $U_G$ it is clear that $U_G$ is real. Covariance
under conventional time reversal symmetry,
\begin{equation} \label{cv}
KU_GK^{-1}=U_G^\dagger
\end{equation}
thus is fulfilled if $U_G$ is symmetric, $U_G=U_G^T$, where $U_G^T$ denotes
the transposed matrix. Using the 
commutation relations between $H,O$ and $D$ (or by evaluating the matrix
numerically, see fig.\ref{fig.syG}) one convinces oneself that $U_G$ is {\em
almost} symmetric. 

\noindent
\begin{minipage}{3.38truein}
\begin{center}
\begin{figure}[h]
\epsfig{file=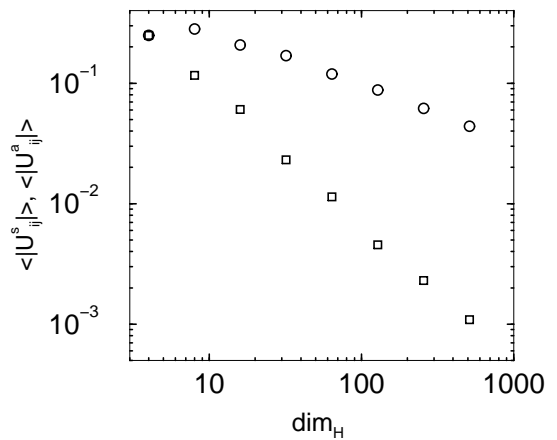,width=8cm} \\[0.2cm]
\caption{Average absolute value of the matrix elements
of the symmetric part of  $U_G$ (circles) and the antisymmetric part
(squares). The former  decays like $1/\sqrt{N}$, the latter like
$1/N$ where $N={\rm dim_H}$ is the dimension of the Hilbert
space.}\label{fig.syG}  
\end{figure}
\end{center}
\end{minipage}
\vspace*{0.1cm}

\noindent
\begin{minipage}{3.38truein}
\begin{center}
\begin{figure}[h]
\epsfig{file=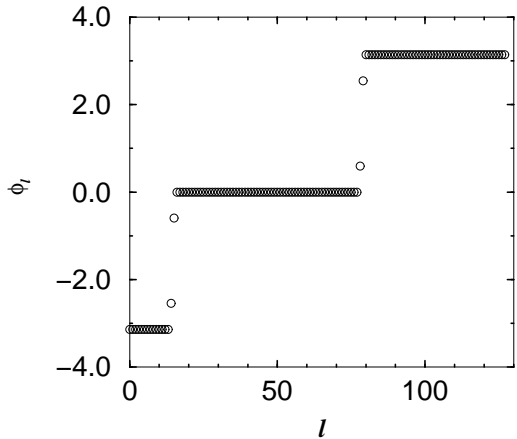,width=8cm} \\[0.2cm]
\caption{Eigenphase spectrum for the Grover algorithm for $q=7$ qubits,
$\xi=64$. The spectrum is symmetric and $-\pi$ and $\pi$ should be
identified. Only two independent eigenvalues differ from $\pm
1$.}\label{fig.ewG}  
\end{figure}
\end{center}
\end{minipage}
\vspace*{0.1cm}

In fact the average absolute value of the matrix elements
of the symmetric part of  $U_G$ decays like $1/\sqrt{N}$, whereas the average
absolute value of the antisymmetric part decays like $1/N$. Just one element
escapes from the general rule set by the average: $U_{G\xi 0}$, the
element in the zeroth column pertaining to the searched element is of order
unity minus a $1/\sqrt{N}$ correction (that is how the whole algorithm
works), whereas $U_{G0\xi}$ plays no such 
special role.  Nevertheless, for $N\to\infty$ the weight of this particular
element is zero and therefore $U_G$ has the unusual property that it is at
the same time unitary as well as real and (almost) symmetric. Thus, all
eigenvalues 
have to be  at the same time on the unit circle, and have to be
real. Therefore all eigenvalues are expected to be 
either 
unity or minus unity! This is well confirmed numerically (see
FIG.\ref{fig.ewG}). In fact it turns out that all eigenvalues, even those
which are not plus or minus one, are to very good approximation 6th roots of
one, and Grover's algorithm is therefore approximately a 6th root of the
unit operation!

The resulting high degeneracy of the eigenphases is in strong
contrast to the level repulsion that goes along with quantum
chaos. However, absence of level repulsion in $U$ does not exclude quantum
chaos.  Indeed, suppose
some $U$ did show level repulsion, then $U^M$ with $M\gg 1$ will in general
not, since the spectrum gets completely mixed by winding it $M$ times around
the unit circle. A Poissonian statistics will be the
consequence. Nevertheless, if $U$ has a chaotic classical counterpart, so
will $U^M$, and to conclude that $U^M$ is {\rm not} chaotic just from the level
statistics is therefore not possible. If the
phases of $U$ are commensurate one can even find an $M$ so that all
eigenvalues are degenerate.\\
\noindent
\begin{minipage}{3.38truein}
\begin{center}
\begin{figure}[h]
\epsfig{file=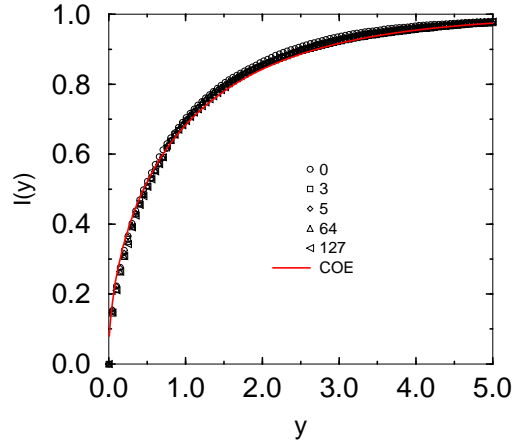,width=8cm} \\[0.2cm]
\caption{Integrated eigenvector distribution $I(y)=\int_0^yR(x)\,dx$ for the
Grover algorithm for 
$n=7$ and various values of 
$\xi$. The different symbols denote different selected numbers
($\xi=0,3,5,64$ and 127). The full line is the integrated Porter Thomas
distribution.}\label{fig.evG}  
\end{figure}
\end{center}
\end{minipage}
\vspace*{0.1cm}

Let us look at the eigenvector statistics of $U_G$. FIG.\ref{fig.evG} shows
that is obeys almost perfectly RMT! 
This is, however, a pure consequence of
the high degeneracy. Indeed, any linear combination of eigenstates in the
subspace pertaining to the degenerate eigenvalue $\lambda=1$ is again an
eigenstate to the same eigenvalue (and correspondingly for
$\lambda=-1$). Thus, a set of eigenvectors can be oriented in a
completely  arbitrary way in this subspace with the only restriction that
they be all normalized and mutually orthonormal. This is just as in RMT,
where the eigenvectors are all statistically independent from the 
eigenvalues, and their orientation is statistically uniform on the whole
possible   hyper-sphere. In the present situation the diagonalization routine
picks initial orientations at random and thus mimics perfect
RMT behavior in both of the two subspaces.

Given the unclear picture presented by the eigenvalue and eigenvector
statistics, it seems reasonable to examine directly the sensitivity with
respect to slight variations in $U_G$. One might do so by applying 
Peres' original scheme, i.e.~creating one perturbed algorithm $U_G'$ and
studying the decay of overlap between $U_G^n|\psi\rangle$ and
$(U_G')^n|\psi\rangle$ as function of $n$ for some random initial
$|\psi\rangle$. 
A natural
parameter in $U_G$ that might be varied is the number of iterations $p$ of
the transformation $DO$. For $p\sim \sqrt{N}$ going from $p$ to $p+1$
appears indeed as a small perturbation. However, all that has been said
about the spectrum of $U_G$ applies to the such perturbed $U_G'$ as well,
i.e.~all eigenvalues will generically be $\pm 1$. A spectral decomposition of
$U_G=\sum_i|u_i^{(+)}\rangle\langle
u_i^{(+)}|-\sum_l|u_l^{(-)}\rangle\langle
u_l^{(-)}|+\sum_k\re^{\ri\varphi_k}|u_k\rangle\langle u_k|$ (where
$\varphi_k$ are the phases different from $0$ and $\pm\pi$) and
correspondingly for $U_G'$ shows that for even $n$ the overlap $\langle
\psi|(U_G^\dagger)^n (U_G')^n|\psi\rangle$ depends on $n$ only due to the
exceptional eigenvalues that differ from $\pm 1$. All the other eigenvalues
contribute only two different values to the overlaps, depending on whether $n$
is even or not. Therefore, varying $p$ (or introducing any other
perturbation that does not lift the degeneracy of the spectrum) does not
lead to exponentially decaying overlap. 

I have therefore applied another perturbation:
All $DO$ factors in $U_G$ are multiplied with a random matrix $V_i$ (drawn
independently for each factor) close to unity,
\begin{equation} \label{}
U_G'=(DO)V_1\ldots(DO)V_pH\,.
\end{equation}
I constructed all of these random
matrices $V_i$ as tensor products of $2\times 2$ orthogonal
matrices $O(2,\varphi)$ close to ${\rm diag}(1,1)$ in the Hilbert space of
each qubit, $V_i=O(2,\varphi_{1i})\otimes\ldots\
\otimes O(2,\varphi_{pi})$, where $\varphi_{1i},\ldots,\varphi_{pi}$ are chosen
randomly and 
independently  from a uniform distribution in the interval
$-\epsilon/2\ldots \epsilon/2$, and $O(2,\varphi)$ is a $2\times 2$
orthogonal transformation acting on a single qubit as  
\begin{equation} \label{ri}
O(2,\varphi)=\left(\begin{array}{cc}
\cos\varphi & \sin\varphi\\
-\sin\varphi & \cos\varphi
\end{array}
\right)\,.
\end{equation} 
Fig.\ref{fig.Groverolap} shows quasiperiodic overlap with almost perfect
revival within 500 iterations for a 5 qubit Grover algorithm
with $\xi=2$ as selected index and $\epsilon=0.1$.  For iteration numbers
exceeding 5000 one notices a slight 
decay, but nevertheless the Fourier spectrum of the decay shows a few very
sharp and strong peaks, indicating the quasi-periodic nature of the
function. We conclude that according to Peres' criterion, Grover's search
algorithm is free of quantum chaos.
 
One might object that as a quantum algorithm, $U_G$ will typically not be
iterated, though. I therefore also applied another criterion proposed
by Schack et al. \cite{Schack94}, which seems to be  more appropriate in the
current situation. These authors examined the distribution of
angles between 
vectors propagated by slightly perturbed unitary matrices, and
embedded Peres' original sensitivity criterion into an information 
theoretical framework. They showed for the example of a kicked top that in
the case of a chaotic quantum map 
the distribution $P(\alpha)$ of angles $\alpha$ between
Hilbert space vectors propagated by many slightly and randomly perturbed
unitary transformations corresponds to that of randomly chosen vectors,
which resembles a Gaussian peak.  An
overall average angle can be steered with a deterministic part of the
random vectors and adapted to $P(\alpha)$ of the Hilbert space vectors. On
the other hand, for integrable quantum 
maps the 
random perturbations lead to many more or less degenerate angles, as the
propagated 
vectors do not explore all Hilbert space dimensions. The angle
distribution therefore typically contains several more or less pronounced
peaks. 
\noindent
\begin{minipage}{3.38truein}
\begin{center}
\begin{figure}[h]
\epsfig{file=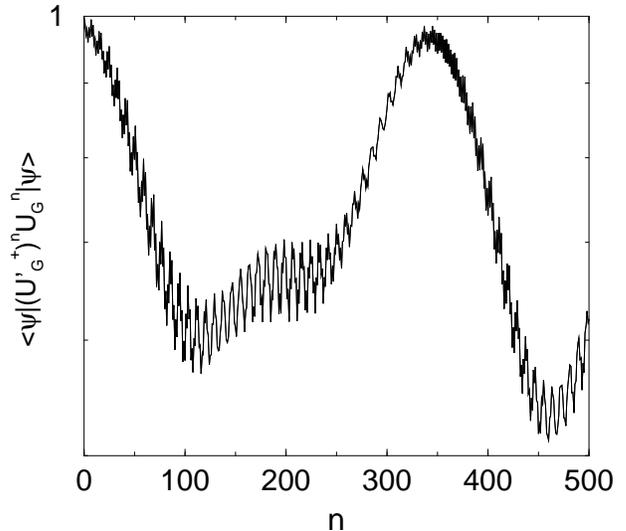,angle=270, width=8cm} \\[0.2cm]
\caption{The overlap between a state propagated by Grover's algorithm and by a
slightly perturbed Grover algorithm ($q=5$ bits, $\xi=2$, $\epsilon=0.1$)
plotted on a logarithmic scale. The
overlap shows quasiperiodic oscillations, characteristic of integrable
behavior. }\label{fig.Groverolap}  
\end{figure}
\end{center}
\end{minipage}
\vspace*{0.1cm}

Instead of adapting the deterministic part of the random vectors, I chose to
``unfold'' the angles,
i.e.~to rescale them by the average angle. It turned out that
``universal'' distributions are obtained by this procedure, both for the
randomly 
drawn vectors as well as those propagated by $U_G'$.  In the latter case one
obtains a distribution which over several orders of magnitude of the
perturbation strength $\epsilon$ is independent of $\epsilon$, \footnote{This
is not true, though, for the perturbed Grover algorithm using the 'digital'
form of perturbation; see below.} in the former
case the unfolded angle distribution is independent of the deterministic
part of the random vectors. 

I have constructed two particular classes of perturbations. The first one
closely corresponds to the original 
recipe by Schack et al. \cite{Schack94}. The perturbed algorithm is
obtained  by multiplying
all $DO$ factors in $U_G$ with only one out of two random orthogonal
matrices close to unity, namely $V_+=O(2,\varphi_1)\otimes\ldots\
,O(2,\varphi_p)$ 
or $V_-=V_+^{-1}$, where again $\varphi_1,\ldots,\varphi_p$ are chosen
randomly and 
independently  from a uniform distribution in the interval
$-\epsilon/2\ldots \epsilon/2$, but are kept fixed for all factors; i.e.~we
obtain $2^p$ perturbed Grover algorithms $DOV_+DOV_+\ldots DOV_+$, $DOV_+DOV_+\ldots DOV_-$, $\ldots$,
$DOV_-DOV_-\ldots DOV_-$. 
A random initial vector is then propagated
by these $2^p$ perturbed matrices and the angle distribution between the
resulting vectors is analyzed. 

Fig. \ref{fig.pofphiG} shows that this sort of 'digital' perturbation does not
entirely randomize the propagated vector. Rather the distribution of angles
shows many pronounced  peaks which means that different perturbations lead
to similar Hilbert space vectors. Thus, from this plot one would conclude
that the behavior is integrable. 
\noindent
\begin{minipage}{3.38truein}
\begin{center}
\begin{figure}[h]
\epsfig{file=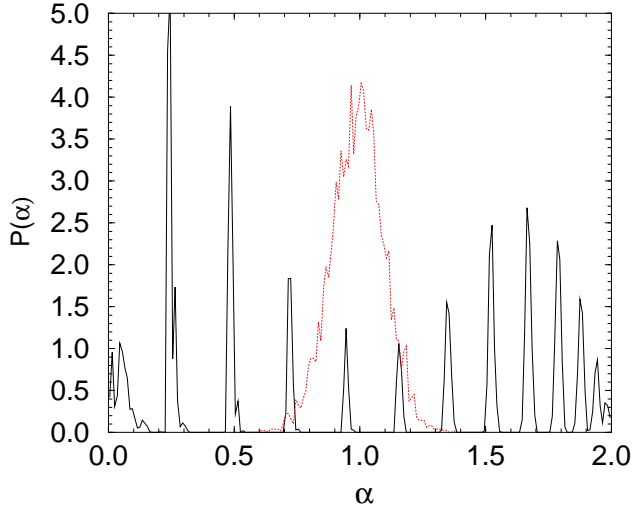,width=7cm,angle=270} \\[0.2cm]
\caption{The distribution of angles between Hilbert space vectors propagated
with $2^p$ perturbed Grover algorithms ($q=5$, $\xi=2$, $\epsilon=0.1$ 
full line) 
is compared to the 
distribution of angles between random vectors 
(dashed line). In both cases the angles were ``unfolded'' such that the
average angles equal unity.}\label{fig.pofphiG}  
\end{figure}
\end{center}
\end{minipage}
\vspace*{0.1cm}
\noindent
\begin{minipage}{3.38truein}
\begin{center}
\begin{figure}[h]
\epsfig{file=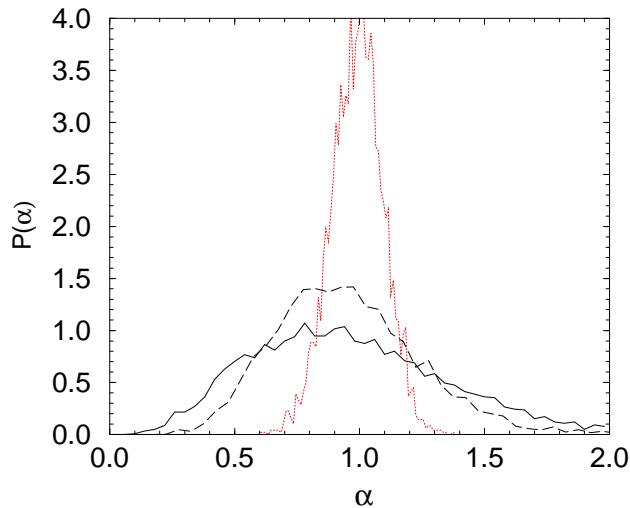,width=7cm,angle=270} \\[0.2cm]
\caption{The distribution of angles between Hilbert space vectors propagated
with 50 perturbed Grover algorithms ($q=5$, $\xi=2$, $\epsilon=0.01$ ;
independent random 
matrices for each 
factor $DO$, full line) is compared to the
distribution of angles between random vectors (narrow peak, dotted
line). All angles were 
``unfolded'' 
such that the average angle is unity. Also shown is the corresponding
distribution for the QFT ($q=5$, $\epsilon=0.01$, dashed
line).}\label{fig.pofphiG2}   
\end{figure}
\end{center}
\end{minipage}
\vspace*{0.1cm}

However, the situation is very different for the second form of
perturbation which corresponds to the one introduced for studying the
decaying overlap, i.e.
instead of choosing between just two matrices $V_-$ and $V_+$ each factor
$DO$ is multiplied with an independent random matrix of the form
$V(\varphi_{1i},\ldots,\varphi_{pi})=O(2,\varphi_{1i})\otimes\ldots\,\otimes
O(2,\varphi_{pi})$. Fig.~\ref{fig.pofphiG2} shows that in this case a broad, 
Gaussian  
like distribution without much further structure arises, much broader in fact
(after unfolding) than the angle distribution obtained from the
random vectors.   

 We may therefore conclude
that for certain perturbations and according to the criterion by Schack et
al.~Grover's search algorithm 
does show hypersensitivity with respect to perturbations. This is quite
surprising in the light of Grover's earlier 
finding that almost any unitary transformation substituted for the Hadamard
matrix still leads to a functioning search algorithm
\cite{Grover98}, and it seems, indeed, that the criterion by Schack et
al. is singled out compared to the other criteria examined.  Note, however,
that there 
is not necessarily a contradiction to Grover's finding, since
for Grover's algorithm to work it is 
enough that only 
the first column of $U_G$ be largely unaffected by the
perturbations (and the same argument applies to all quantum algorithms that
start from just one initial state, typically the state $|00\ldots 0\rangle$)!
The present result is more general in as much as it makes a
statement about the sensitivity of the entire matrix with respect to
perturbations.  

How rapidly does the average fidelity decrease with the perturbation
strength? If we measure lack of fidelity as average absolute error of all
matrix 
elements, the answer is: linearly, over several orders of magnitude of
$\epsilon$ ($0.0001\le\epsilon\le 0.1$), as can be
seen from Fig.\ref{fig.errofeps}, and as would have been expected from
perturbation theory. 

Let me now come to the quantum Fourier transform (QFT). Since it is a
universal part of several proposed quantum 
algorithms including Shor's algorithm  \cite{Shor94}, it makes sense to give
the QFT special attention. 
The quantum Fourier transform $U_{FT}$ on a $n$ bit register (with qubits
indexed as
$0\ldots n-1$) can be
constructed from one-- and two--qubit operations as \cite{Shor94}
\begin{eqnarray} \label{shorc}
U_{\rm FT}&=&FH_0S_{0,1}S_{0,2}\ldots
S_{0,n-1}H_1\ldots H_{n-3}S_{n-3,n-2}\nonumber\\
&&S_{n-3,n-1}H_{n-2}S_{n-2,n-1}H_{n-1}\,, 
\end{eqnarray}
where $S_{j,k}$ is a conditional phase shift matrix between qubits $j,k$
defined by $S_{j,k}={\rm diag}(1,1,1,\exp(\ri\pi/2^{k-j}))$ in the basis $00$,
$01$, $10$, and $11$ formed by the two qubits $j$ and $k$. The matrix
$F$ flips all qubits. With all the
two--qubit interactions introduced, one would naively expect quantum
chaos. However, the QFT is constructed such that in the whole Hilbert space
$U_{\rm FT}$ is very simple and symmetric,
\begin{equation} \label{QFT}
U_{{\rm FT}lk}=\frac{1}{\sqrt{N}}\re^{\ri 2\pi lk/N}\,.
\end{equation}  
One easily convinces oneself that $U^4={\bf 1}$! Thus, all
possible eigenphases are  $0$, $\pi$, and $\pm\pi/2$. The situation is
therefore even simpler than in Grover's algorithm: There is an exact
relation that dictates a high degeneracy of only four possible eigenphases. So
again, there is no level repulsion.   
\noindent
\begin{minipage}{3.38truein}
\begin{center}
\begin{figure}[h]
\epsfig{file=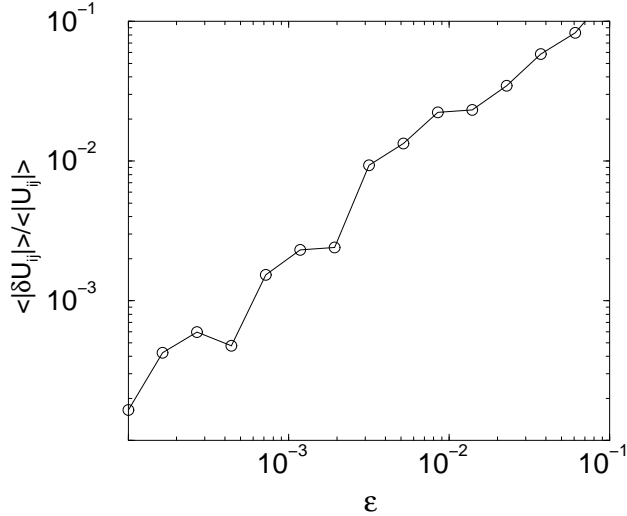,width=7cm,angle=270} \\[0.2cm]
\caption{The absolute average error of the matrix elements of Grover's
search algorithm $U_G$ (in units of the absolute average matrix element) for
random perturbations (independent matrices $V(\varphi)$ for each
factor $DO$); $q=5$, $\alpha=2$.}\label{fig.errofeps}  
\end{figure}
\end{center}
\end{minipage}
\vspace*{0.1cm}

The matrix $U_{\rm FT}$ is covariant under
conventional time--reversal: $KU_{\rm FT}K^{-1}=U^*=U^\dagger$ since
$U=U^T$. 
And a numerical  evaluation of the eigenvector statistics leads again to
good agreement with the Porter--Thomas distribution, which, however, results
once more from the high degeneracy of the eigenvalues and not from quantum
chaos. 
\noindent
\begin{minipage}{3.38truein}
\begin{center}
\begin{figure}[h]
\epsfig{file=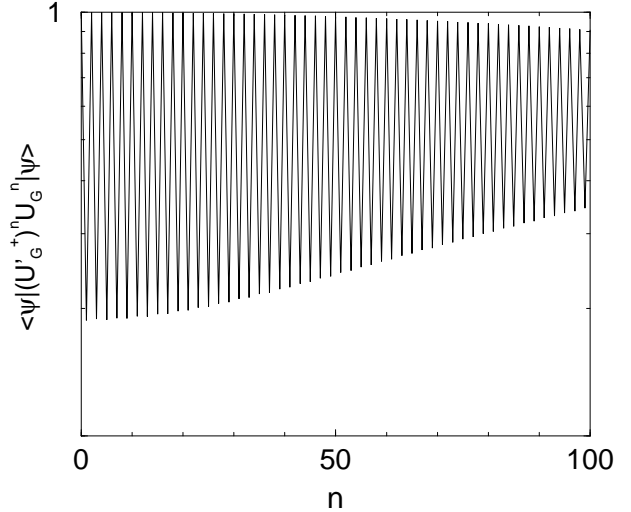,angle=270, width=8cm} \\[0.2cm]
\caption{The overlap between a state propagated by the QFT and by a
slightly perturbed QFT ($q=5$ bits, $\epsilon=0.1$) plotted on a logarithmic
scale. The
overlap shows quasiperiodic oscillations, characteristic of integrable
behavior. }\label{fig.qftolap}  
\end{figure}
\end{center}
\end{minipage}
\vspace*{0.1cm}

For studying the sensitivity of $U_{\rm FT}$ with respect to perturbations,
I slightly perturbed all phases in the conditional phase gates by adding an
additional random phase with the
same (on the average) relative amount for all gates
\cite{relphase}. Fig.~\ref{fig.qftolap} shows that the overlap of a random
state propagated with a slightly perturbed matrix $U_{\rm FT}'$ and the same
state propagated with the original $U_{\rm FT}$ shows again quasiperiodic
oscillations. The basic period is two, as could be expected from $(U_{\rm
FT}^2)_{ij}=\delta_{i,N-j}$, i.e.~the exact QFT leads back to the same
starting vector up to a relabeling of the indices when applied
twice. The perturbed QFT only leads to a small deviation from this relabeled
initial vector, and therefore almost perfect overlap is  restored every
second iteration. Thus, according to Peres' criterion the QFT is free of
quantum chaos. 

Fig.~\ref{fig.pofphiG2} shows the angle
distribution resulting from propagating a random initial vector by 100
differently perturbed matrices for an algorithm running on five
qubits. The Hilbert space vectors are complex now, so I defined the angles
$\alpha_{i,k}$ as 
 $\alpha_{i,k}=\arccos(|\langle \psi_i|\psi_k\rangle|/\sqrt{\langle
\psi_i|\psi_i\rangle\langle \psi_k|\psi_k\rangle})$, and all angles were
again rescaled according
to $\alpha_{i,k}\longrightarrow \alpha_{i,k}/\langle \alpha \rangle$, where
$\langle \alpha \rangle$ is the average angle for all pairs $i\ne k$. The
resulting 
distribution depends for a small number $q$ of qubits still substantially on
$q$. For $q=3$ a distribution is obtained that is remarkably close to the
famous Wigner Dyson surmise for the orthogonal ensemble (\ref{WD}), but for
larger numbers of qubits the distribution approaches more and more a
Gaussian, not too different from the distribution obtained from the Grover
algorithm (see Fig.~\ref{fig.pofphiG2}). Again the distribution is
stable in the range I examined it, 
$0.0001\le \epsilon\le 0.1$ .  Thus, also the
QFT shows hypersensitivity with respect
to random perturbations as judged by the criterion of Schack et al., and
in contrast to Peres' original criterion!

On the other hand it is known that if the controlled phase gates are
performed to a precision $\Delta=1/p(n)$ where $p$ is a polynomial in the
number of qubits, then the maximum error of the final state
$U_{QFT}|\psi\rangle$ for all input states $|\psi\rangle$ is of order
$n^2/p(n)$ \cite{Nielsen00}.  Thus, only polynomial precision is needed
for the controlled phase gates. Coppersmith has shown, indeed, that an
approximate QFT can be obtained by dropping the
gates with the exponentially  small phases altogether
\cite{Coppersmith94}.  

{\em In summary} I have shown that both Grover's search algorithm and the
QFT give rise to the same  unusual combination of quantum signatures of
chaos and of integrability. Strong symmetries lead to a
large degeneracy of the spectra of eigenvalues of the unitary matrices
representing these algorithms. In fact, the QFT is a fourth
root of the unity matrix, and Grover's algorithm is to a good approximation
a 6th root of unity! The corresponding lack of level repulsion
would be commonly interpreted as absence of quantum chaos. The eigenvector
statistics closely 
follows RMT predictions, and one would commonly interprete this as a
signature of quantum chaos. However, as shown above, it is here but an
artefact arising from the highly degenerate spectrum. The overlap between a
random state propagated by a 
perturbed algorithm and the same state propagated by the
corresponding unperturbed algorithm shows quasiperiodic oscillations both
for Grover's algorithm and the QFT, thus signaling absence of quantum
chaos, in agreement with earlier studies addressing the stability of these
codes. The 
 only criterion indicating quantum chaos and not evidently explainable by an
artefact 
is the distribution of 
angles between vectors 
propagated once by many slightly disturbed algorithms. After unfolding the
angles 
universal distributions are obtained for large enough Hilbert spaces, both
for Grover's algorithm and 
for the QFT,  that
resemble the one for 
random vectors. 

{\em Acknowledgment:}
I would like to thank Henning Schomerus for a useful
discussion. This work was 
supported by the Sonderforschungs\-be\-reich 237 ``Unordnung und 
gro{\ss}e Fluktuationen".

\end{multicols}

\end{document}